\documentclass{aa}
\usepackage{graphicx}
\usepackage{psfig}
\begin{document}
   \title{First SIMBA observations toward CH$_{3}$OH masers}

   \subtitle{Masers from blind surveys mark the earliest stages of massive star formation}

   \author{M. Pestalozzi
          \inst{1}
          \and
          E.M.L. Humphreys
          \inst{1}
          \and
          R.S. Booth
         \inst{1}
          }

   \offprints{michele@oso.chalmers.se}

   \institute{Onsala Space Observatory, 
               S-43992 Onsala, Sweden\\
               email:michele@oso.chalmers.se 
             }

   \date{Received December 2001 / Accepted January 2002}

   \abstract{We report SIMBA 1.2\,mm dust continuum observations of the
        environments of eight methanol maser sources, all discovered during
        spatially fully-sampled, untargeted  
        surveys of the galactic plane. We summarise our search for possible
        associations of the masers with IR sources (IRAS and MSX) and find
        that it is not always possible to make definite associations. A
        preliminary characterisation of the IR sources found in the maser
        neighbourhood is given according their position in the
        [60-25] -- [25-12] colour-colour diagram.
   }

   \maketitle
%


\section{Introduction}

Massive star formation has attracted increasing interest in recent years, with 
the mounting evidence that methanol masers provide excellent tracers and
probes of the high-mass star formation process (e.g. Hunter et al. 1998;
Minier et al. 2001). Such evidence has led to several surveys aimed at
identifying new high-mass stars, often targeting IRAS colour-selected UCHII
regions (Caswell et al. 1995; Schutte et al. 1993; Szymczak et
al. 2000). Although making significant contributions to the discovery of
new massive star forming regions (MSFRs), such surveys have yielded only a
 $\sim$15\% detection rate of methanol masers.  

An alternative approach for discovering MSFRs has been provided by {\it blind}
surveys of 6.7\,GHz methanol masers. In particular, the surveys of Ellingsen
et al. 
(1996) and Pestalozzi et al. (2002, in prep.) have sampled in a consistent
manner  
large regions of the galactic plane from the southern and northern hemispheres
respectively. Since the majority of methanol masers revealed in these surveys
does not have a clear association (within the positional accuracy) with IRAS
sources, the question has arisen: could they be tracing early stages of the 
 high-mass star formation process, in which the still highly-embedded, nascent 
star is radiating much of its flux in the millimetre and submillmetre regions
(as suggested by Walsh et al. 1999)?

With the aim of answering this question, in addition to ascertaining the SEDs 
for regions for which there is existing IR data (IRAS and MSX), we report here
on 1.2\,mm bolometer observations of the regions surrounding eight 6.7\,GHz
methanol masers detected in blind surveys. This constitutes the first part of
our follow-up investigations for understanding the methanol maser phenomenon,
and the evolutionary route to forming massive stars. 

\vskip -1.5cm



\section{Observations}

Observations were made using the SIMBA 1.2\,mm bolometer array
on the SEST on 19 October 2001. SIMBA is a 37-channel hexagonal array 
in which the HPBW of a single element is about 24" and  the separation between 
elements on the sky is 44". The bandwidth in each channel is about 
50\,GHz. Spectral line emission present in the band may contribute to the
total continuum flux values reported in this paper, though not more than 30\%
as in the case of molecular rich clouds (e.g. IRC+10216, L-\AA{} Nyman,
priv. comm.). These observations were made in 
fast-mapping mode. Areas typically of 600" x  384" were imaged in order to
detect nearby IRAS sources in the field, using a scan speed of
80"s$^{-1}$. Typical total observation times were around one
hour per source, during which 5 to 6 maps were produced. Zenith opacities
lay in the range 0.217 -- 0.270 
throughout the observing run. The resulting data were reduced with the MOPSI
mapping software package developed by R. Zylka, using a "deconvolution"
algorithm\footnote{Developed within the SIMBA collaboration} to remove 
the contribution of the electronics arising from 
the fast-mapping observing mode, the "converting" 
algorithm (Salter 1983) to convert the coordinates from
rectangular to equatorial, and partly the NOD2 and GILDAS
libraries. The calibration of our data was performed using 1.2\,mm Uranus data
 for 19 October. The multiplication factor between
counts and Jy was 166.5 mJy count$^{-1}$ beam$^{-1}$ (Table
\ref{tab:soulist}).


\begin{table*}[t]
\begin{center}
\vspace{0.5cm}
\begin{tabular}{lllcccc}
\hline \hline
Source & RA(2000) & Dec(2000) & v$_{lsr}$ [kms$^{-1}$] & Peak [Jy] &
Int. flux [Jy/beam] & rms \\
\hline
G40.25$-$0.19$^1$ & 19 05 32.6 & +06 25 38 & +70.0 & 2.55 & 2.64 & 0.164 \\
G41.34$-$0.14$^1$ & 19 07 21.870 & +07 25 17.34 & +14.0 & 0.30 & 0.38 & 0.036 \\
327.12+0.51$^{2,3,4,5}$ & 15 47 32.729 & $-$53 52 38.90 & -87.1 & 1.54 & 2.06 & 0.070 \\
327.59$-$0.09$^3$ & 15 52 36.824 & $-$54 03 18.97 & -86.3  & 0.34 & 0.44 & 0.050 \\
327.62$-$0.11$^3$ & 15 52 50.241 & $-$54 03 00.71 & -97.5 & 0.43 & 0.72 & 0.050 \\
329.34+0.15$^{3,4}$ & 16 00 33.154 & $-$52 44 40.00 & -106.5 & 4.37 & 6.67 & 0.132 \\
332.35$-$0.44$^3$ & 16 17 31.560 & $-$51 08 21.55 & -53.1 & 0.67 & 0.76 & 0.052 \\
333.13$-$0.56$^3$ & 16 21 35.742 & $-$50 40 51.29 &  -56.8 & 3.06 & 4.83 & 0.200 \\
\hline
\end{tabular}
\end{center}
\caption{\label{tab:soulist} 
Source list with J2000 coordinates, radial velocities of the methanol
maser emission, 1.2\,mm peak, integrated flux emission and rms. References:
1) Pestalozzi et al. 2002 in prep.; 2) MacLeod \& Gaylard 1992; 3) Ellingsen et al. 1996;
4) Walsh et al. 1998; 5) Phillips 1998.}
\end{table*}

\begin{figure}
\centerline{
\psfig{file=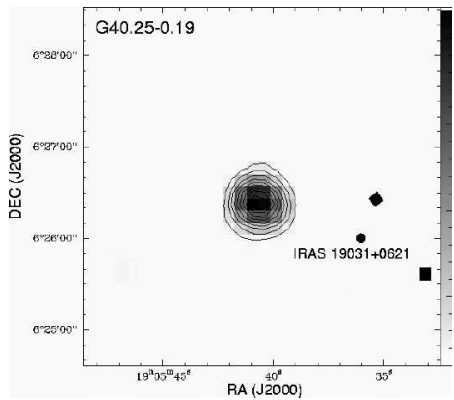,height=52mm}
}
\caption{\label{fig:33315} This figure and Fig. \ref{fig:all_sou} show SIMBA
  maps of the 1.2\,mm dust continuum toward 
  the methanol masers listed in Table \ref{tab:soulist}. Circles indicate the
  IRAS source, diamonds the MSX source, squares the methanol maser
  position. Contours and grey scale indicate the 1.2\,mm intensity in Jy. Here
  we show the region around G40.25-0.19.} 
\end{figure}

\section{Results \& Discussion}

Figs.\ref{fig:33315} and \ref{fig:all_sou} show the images obtained at 
1.2\,mm for the sources listed in Table \ref{tab:soulist}. The same table
gives the peak and integrated fluxes of the detections of the 1.2\,mm dust
continuum. 
For the position accuracy of the 1.2\,mm emission we adopt the FWHP of one
channel of the array (24"). For the IR sources we refer to the literature
(1-2' for IRAS, 5" for MSX). The positions of the southern methanol masers are 
accurate to better than 1 arcsec (Ellingsen, priv. comm.). The
position of 
G41.34-0.14 has been determined to an accuracy of 5 mas\footnote{Observed
  with the Jodrell-Bank $-$ Cambridge baseline}, while for
G40.25-0.19 the uncertainty is larger (about 1'). We refer to
Ivezi\'c \& Elitzur (2000) for the discussion about the IRAS sources found
close to our methanol masers. In their reanalysis of the colour 
properties of the galactic IRAS sources they find a consistent separation of
the Point Source Catalogue (PSC) into four classes that, according to the
authors, are due to different 
spatial distribution of the dust around those sources. Confirmation comes from
radiative transfer calculations for a point source surrounded by a dust
envelope. By varying $p$ in the power law density profile $r^{-p}$ the authors
produce solutions characterised by the optical depth $\tau_{\nu}$. Making use
of that study we are able to infer density profile and optical depth of
some of our sources.

\subsection{G40.25-0.19}

G40.25-0.19 has been detected both by the Onsala blind survey and by the
colour-selected survey of Szymczak et al. (2000). Our SIMBA map of G40.25-0.19
shows only one site of emission which has no associated IR emission. The
methanol maser emission is well-separated from the 1.2\,mm 
emission peak. Poor positional accuracy of the methanol maser does
not allow us to draw conclusions about associations. Further continuum and
molecular line observations are required in order to characterise the nature
of this region.

\subsection{G41.34-0.14}

\begin{centering}
\begin{figure*}
\centerline{
\psfig{file=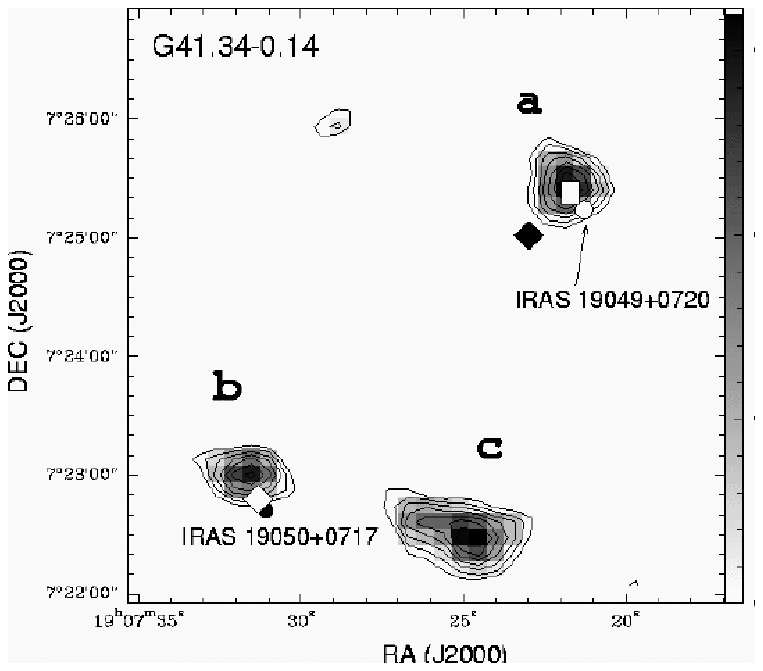,height=50mm}
\hspace{0.1cm} \psfig{file=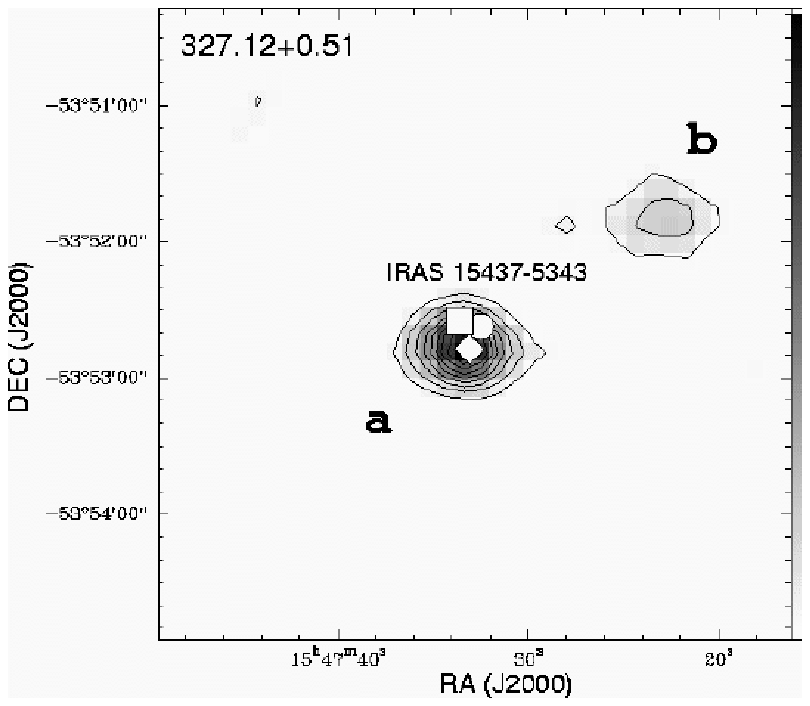,height=50mm}
}
\vspace{0.3cm}
\centerline{
\psfig{file=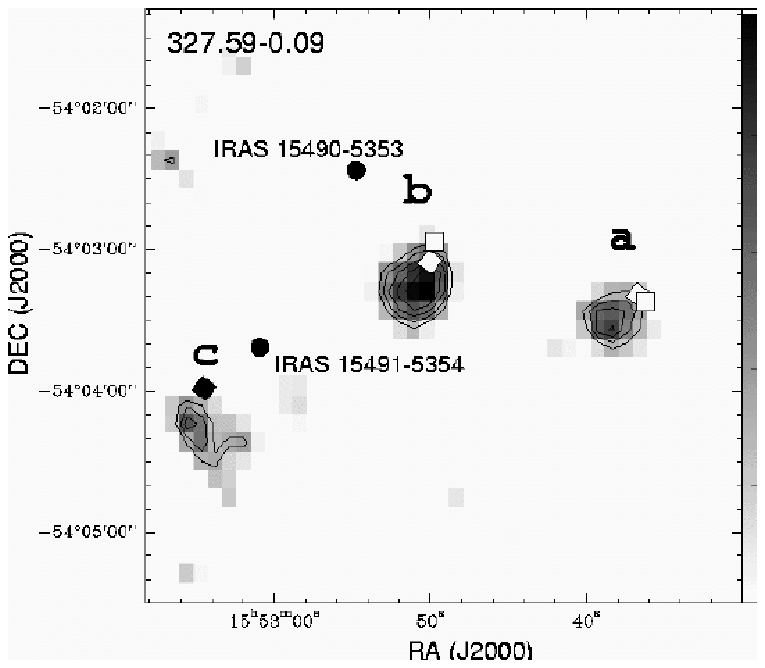,height=50mm}
\hspace{0.1cm} \psfig{file=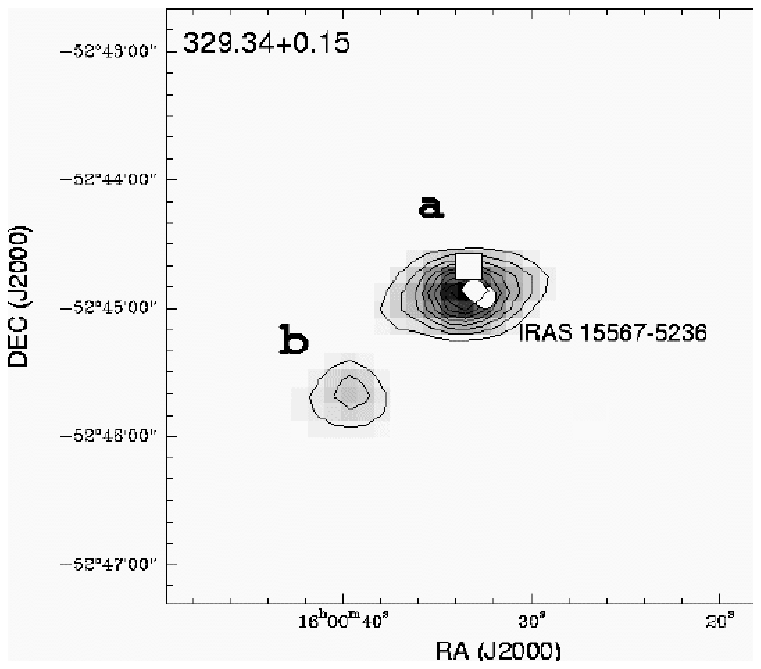,height=50mm}
}
\centerline{\psfig{file=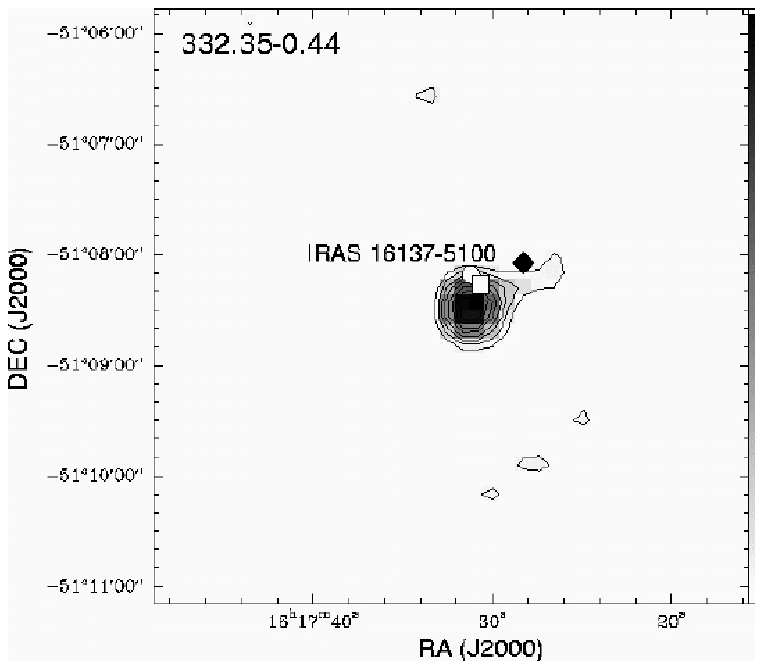,height=50mm}
\hspace{0.1cm} \psfig{file=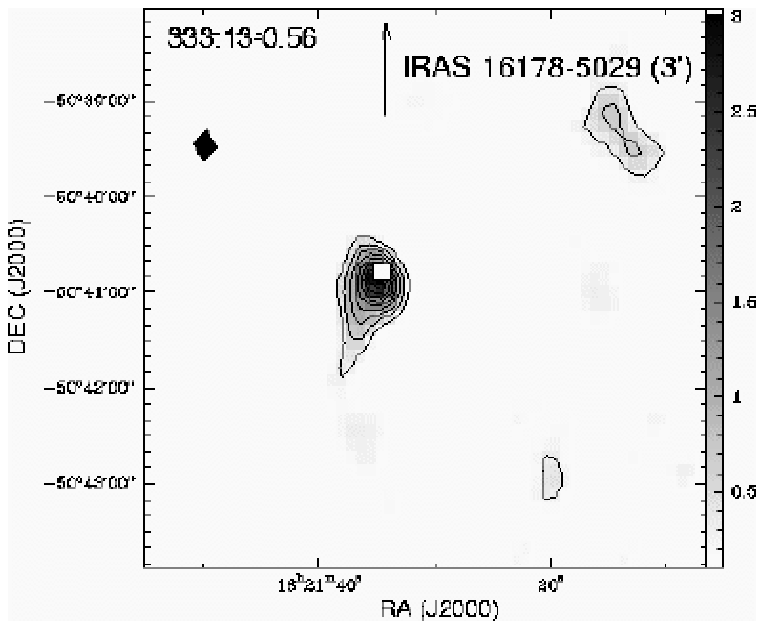,height=50mm}
}

\caption{\label{fig:all_sou} See Fig. \ref{fig:33315}.}
\end{figure*}
\end{centering}

G41.34-0.14 (component \verb*|a| in Fig. \ref{fig:all_sou}) was detected by
Pestalozzi et al. (2002) during  the  Onsala 6.7\,GHz blind survey. It has two
main features, the strongest is approximately 20 Jy. The
associated IRAS source has been systematically discarded by colour-selected
surveys, since it does not match any criteria for UCHII regions. It seems to
have a flat density profile ($p=0$) and an optical depth of $\tau_{\nu}\sim
0.1$. In the same field we have discovered two other nearby 1.2\,mm sources,
of which \verb*|b| has both an IRAS and MSX counterpart, whilst \verb*|c| does
not have any IR counterpart.

\subsection{327.12+0.51}

This strong (80 Jy, three spectral features) methanol maser was first detected
by MacLeod \& Gaylard (1992). It has OH and H$_2$O 
associated masers (Caswell et al. 1980; Batchelor et al. 1980), making it a
classical high mass 
star formation region. IRAS 15437-5343 shows colours of an UCHII region and
appears deeply embedded (high optical depth $\tau_{\nu}\sim 100$). Its density
profile is fairly flat ($0 \le p \le 0.5$). 8.6\,GHz and 6.67\,GHz continuum
emission (12 and $<$12 mJy respectively) have been observed by Walsh et
al. (1998) at more than 1 arcmin offset from the IRAS 
source. Phillips (1998) detected a 0.5\,mJy 8.6\,GHz continuum peak at 1
arcsec from the methanol maser. VLBI observations of the methanol maser
(Phillips 1998) show a linear distribution of the maser components with a
linear velocity gradient. Assuming a Keplerian disc, the mass for the central
object is estimated to be 14.7 M$_{\odot}$.

\subsection{327.59-0.09 \& 327.62-0.11}

There are two weak (3 Jy), single featured  methanol masers in the SIMBA
field, \verb*|a,b| as 
327.59-0.09 and 327.62-0.11 respectively (Ellingsen et al. 1996). Both show
clearly 
associated 1.2\,mm and MSX counterparts. IRAS 15490-5353 could be related to
component \verb*|b|. No other counterpart is visible. As for 333.13-0.56, we
suggest that the methanol masers trace very deeply embedded objects.

A third 1.2\,mm peak of emission is visible in the field (component
\verb*|c|), having an MSX and probably an IRAS counterpart, but no methanol
maser. 



\subsection{329.34+0.15}

This single featured 15 Jy methanol maser has been discovered by Ellingsen et
al. (1996) and is clearly associated with IRAS 15567-5236, an UCHII
region. Schutte et al. (1993)  
did not detect any methanol maser above their 5 Jy limit. Walsh et al. (1998)
detect both a 8.6 and a 6.67\,GHz continuum peak ($<$20 and 270 mJy/beam 
respectively). The colours of IRAS 15567-5236 suggest that we are observing a
young object with very low optical depth ($\tau_{\nu} < 0.1$). The density
profile is fairly flat, p=0.5.

\subsection{332.35-0.44}

This weak (4 Jy, one spectral feature) methanol maser was discovered by
Ellingsen et al. (1996). It seems to lie on the edge of a 5\,GHz peak of
emission 
(Haynes et al. 1978). Both IR sources appearing in the SIMBA field fall on top
of the 1.2\,mm peak of emission, within the positional accuracies. Since
the angular distance between the IR  sources and the methanol maser is greater
than 1', there is no clear association between the two emitting
components. Ellingsen et al. exclude it as an association because the colours
of 
the IRAS source are not representative for an UCHII region. The IRAS
source seems to have a flat density profile ($p=0$) and low optical depth,
$\tau_{\nu}\sim 0.1$.  

%
%

\subsection{333.13-0.56}

This 6.7\,GHz maser source was detected by Ellingsen et al. (1996). It shows
three 
distinct spectral features, the strongest being about 17 Jy. It does not have
any IRAS counterpart within the positional accuracies. An MSX source lies
at a distance of $>$1.5 arcmin from both the methanol maser position and the
peak in the 1.2\,mm emission. The methanol maser and the strong 1.2\,mm
emission are spatially coincident within the positional
accuracies. This indicates that the methanol maser is tracing a very deeply
embedded object.


\section{Conclusions}

We can divide the observed source list into two separate groups: 333.13-0.56,
327.59-0.09, 327.62-0.11, G41.34-0.14 in the first group; 
332.35-0.44, 327.12+0.51, 329.34+0.15 in the second. G40.25-0.19 is left out
of the discussion because of the poor positional accuracy of the methanol
maser. The first group collects
all sources with either no IRAS counterpart or a weak one. These sources could
have been detected only by blind surveying. The 1.2\,mm dust continuum
confirms the existence of dust where a methanol maser is detected. We consider
them as the group of young, still embedded sources. The sources of the second
group have clear IRAS and MSX couterparts. The colours of the IRAS sources
fulfil the selection criteria by Wood \& Churchwell (1989) or Szymczak et
al. (2000), so we can consider them as (young) UCHII regions. The lack of cm
continuum emission towards most of the sources can be explained in two ways:
we can see it as a further argument to strengthen our hypothesis that these
sources are in early stages of the stellar evolution; it could also be argued
that those central stars are not massive enough to produce Lyman $\alpha$ 
photons. We prefer the former idea, as methanol masers are known to trace
massive stars.

Despite the clear separation into two classes of the sources, it remains
nevertheless difficult to find correlations between the 1.2\,mm flux density
and the methanol flux density. Furthermore it is puzzling to notice that
327.59-0.09 and 327.62-0.11 do have an 1.2\,mm and an MSX
counterpart but no clear IRAS counterpart. These topics will be investigated
further, by observing in the submillimetre range (850-450 $\mu$m).

The observations lead us to the following conclusions:

\begin{itemize}

\item Blind surveys of methanol masers facilitate the detection of 
  young, still deeply embedded high-mass star formation regions;
\item IRAS sources associated with methanol masers show a flat density 
  profile ($0 \le p \le 0.5$), which means that they can be classified as
  objects in an early stage of massive star formation. 

\end{itemize}

\begin{acknowledgements}

      SIMBA was built and installed at the SEST at La Silla (Chile) within an
      international collaboration between the University of Bochum and the
      MPIfR in Bonn, Germany, the Swedish National Facility
      (OSO) and ESO. 
      We thank the SEST team for their help with the observations
      and data reduction. We thank Dr M. Albrecht and collaborators
      for providing the planet maps for our data calibration.
      We thank Dr Simon Ellingsen for providing accurate
      positions of some methanol masers and useful comments on the
      manuscript. EMLH thanks STINT for funding 
      her  
      research position and the Vetenskapr\aa{}det for a travel 
      bursary to the SEST.

\end{acknowledgements}

\end{document}